\documentclass[a4paper,12pt]{article}
\usepackage{amsmath,amsfonts,amssymb,epsfig,subfigure,color}

\newcommand{\be}{\begin{equation}}
\newcommand{\en}{\end{equation}}

\newcommand{\ii}{\textrm{i}}

\renewcommand{\vec}[1]{\boldsymbol{#1}}
\begin{document}
\numberwithin{equation}{section}
%++++++++++++++++++++++++++++++++++++++++++++++++++=++++++++++++++++++
\title{Transverse Waves in Nonlinearly Elastic Solids and the Milne-Pinney (or Ermakov) Equation}
%++++++++++++++++++++++++++++++++++++++++++++++++++++++++++++++++++++
\author{Michel Destrade\\
School of Electrical, Electronic, and Mechanical Engineering,\\ 
University College Dublin, Belfield, Dublin 4, Ireland.\\
\texttt{e-mail: michel.destrade@ucd.ie}\\
\\
Giuseppe Saccomandi\thanks{corresponding author}\\
Dipartimento di Ingegneria Industriale, \\ Universit\'a degli Studi di Perugia, 06125 Perugia, Italy\\
\texttt{e-mail: saccomandi@mec.dii.unipg.it}}
\date{\emph{Dedicated to Michael Carroll \\on the Occasion of his 75th Birthday}}

%++++++++++++++++++++++++++++++++++++++++++++++++++++++++++
\maketitle
%++++++++++++++++++++++++++++++++++++++++++++++++++++++++++
\begin{abstract}
 
We establish a connection between the general equations of nonlinear
elastodynamics and the nonlinear ordinary differential equation of Pinney
[Proc. Amer. Math. Soc. 1 (1950) 681]. 
As a starting point, we use the exact travelling wave solutions of nonlinear elasticity discovered by Carroll [Acta Mechanica 3 (1967) 167]. 
The connection provides a method for finding new exact and approximate dynamic solutions for neo-Hookean and Mooney-Rivlin solids, and for the general third- and fourth-order elasticity models of incompressible solids. 
\end{abstract}

\newpage

%%%%%%%%%%%%%%%%%%%%%%

\section{Introduction}

%%%%%%%%%%%%%%%%%%%%%%

The equations of elastodynamics belong, under the usual constitutive assumptions, to the class of hyperbolic systems. 
It is known that if a hyperbolic system is nonlinear, then smooth solutions of initial-value or initial boundary-value problems do not usually exist globally in time, and that singularities will develop, typically after a finite time, even when the initial or boundary data are smooth \cite{Daf}. 
Existence of smooth solutions to the initial-value problem of nonlinear elastodynamics is possible only in special situations and only for \textit{body waves}. 

To the best of our knowledge, the only explicit examples of smooth solutions to the equations of nonlinear elastodynamics are the \emph{Carroll waves} \cite{Ca0, Ca1, Ca2, Ca3}.  
These solutions were first investigated in \cite{Ca0} in the form of finite-amplitude, circularly-polarized, shear plane waves traveling in \emph{any} equi-biaxially deformed solid, as
\be \label{carroll}
x= \mu_S X + A \cos(k Z - \omega t), \quad 
y = \mu_S Y + A \sin(k Z - \omega t), \quad
z = \lambda_S Z,
\en
where $A$, $k$, $\omega$, $\mu_S$ and $\lambda_S$ are arbitrary constants  (in incompressible materials, $\mu_S^2 \lambda_S=1$). 

Recently these solutions have been generalized by Destrade and Saccomandi \cite{77, 97} to dissipative and dispersive materials and in the case of a time dependent homogeneous underlying motion by Rajagopal \cite{R1} and Pucci and Saccomandi \cite{Pucci}.  

The aim of this note is to show that there exists a nice connection between some of the remarkable solutions of Carroll and the \emph{Milne-Pinney equation} ordinary differential equation (also known as the Ermakov  equation \cite{Leach}). 
This is the nonlinear ordinary differential equation
\be
y'' + p(x)y + cy^{-3} = 0,
\en
for $c$ constant and $p(x)$ given.
In such a way it is possible to uncover some simple new exact solutions to the equations of nonlinear elastodynamics. These new solutions show once again the relevance of the Carroll waves solutions to our understanding of elastodynamics.

 %%%%%%%%%%%%%%%%%%%%%%

\section{Basic Equations}

%%%%%%%%%%%%%%%%%%%%%%

We call $\vec{x}(\vec{X},t)$ the current position of a particle which was located at $\vec{X}$ in the reference configuration. 
We follow Pucci and Saccomandi \cite{Pucci} and consider the motion
\begin{equation}
x=\mu (t)X+f(Z,t),\qquad y=\mu (t)Y+g(Z,t), \qquad z=\lambda (t)Z,  \label{p1}
\end{equation}
which consists of two shearing motions $f$ and $g$ in the $Z$-direction, combined with a \textit{time-dependent} homogeneous biaxial stretch $\lambda(t)$ (here, $\mu \equiv \lambda^{-1/2}$). 
This class of motions has been considered by Rajagopal \cite{R1}; 
it is a generalization of the Carroll waves \eqref{carroll}.

Two kinematic quantities associated with this motion are the deformation gradient and the left Cauchy-Green strain tensor,
\be
\vec{F} = \partial \vec{x} / \partial \vec{X}, \qquad \vec{B} = \vec{FF}^{T},
\label{05}
\en
respectively.
We focus on incompressible solids, which can undergo only isochoric motions, so that $\det \vec{F}=1$ at all times (this is indeed the case for this motion).  

We also require that the material be hyperelastic  and isotropic, and so we introduce the strain energy density  $W = W(I_{1},I_{2})$, where $I_{1}$ and $I_{2}$ are the first and second principal invariants of $\vec{B}$, respectively.  
For isochoric motions, they are given by
\be
I_{1}=\text{tr} \ \vec{B}, \qquad I_{2}=\text{tr}(\vec{B}^{-1}).
\label{0011}
\en
The general representation formula for the Cauchy stress tensor $\vec{T}$ is
\begin{equation}
\vec{T}=-p\vec{I}+2 W_1 \vec{B} - 2 W_2 \vec{B}^{-1},
\label{1}
\end{equation}
where $p$ is the yet indeterminate Lagrange multiplier introduced by the constraint of incompressibility, and $W_1 \equiv \partial W / \partial I_1$, $W_2 \equiv \partial W/\partial I_2$.

Now, the balance { equation} of linear momentum, in the absence of body forces, is
\begin{equation}
\text{div}\ \vec{T} = \rho \ \partial^2\vec{x}/\partial t^2,  \label{2}
\end{equation}
where $\rho$ is the (constant) mass density.

Standard computations \cite{Pucci} lead to the following specialization of the balance equations to the motion \eqref{p1},
\begin{align}
& \rho \left( f_{tt}+\mu_{tt}X \right)  =  - p_x + \left[ 2(\lambda W_1+W_2) f_{Z} \right]_z,
\notag  \\
& 
\rho \left( g_{tt} + \mu_{tt} Y\right)  =  - p_y + \lambda \left[ 2(\lambda W_1+W_2) g_{Z} \right]_z,
\label{p9}
\notag \\
& \rho \lambda _{tt}Z = - p_z + \left[ 2\lambda ^{2}W_{1}-2 \lambda^{-1} \left( \mu^{2}
f_{Z}^{2}+g_{Z}^{2}\right) W_{2} \right]_z,
\end{align}
where letter subscripts denote partial differentiation.
Then we write these equations in the reference configuration, and we introduce the \emph{shear strains} $F \equiv f_Z$ and $G \equiv g_Z$, and the \textit{generalized shear modulus} $Q \equiv 2(W_1+\lambda^{-1}W_2)$. We point out that here 
\be
I_1=2\mu^2+\lambda^2 + F^2 + G^2, \qquad 
I_2=2\lambda^2+\lambda^{-1}(\mu^2+F^2 + G^2).  
\en

Finally we eliminate the Lagrange multiplier $p$, and we obtain the determining equations for the strains $F$ and $G$,
\be  \label{equa}  
\rho \left( F_{tt}-\frac{\mu_{tt}}{\mu}F \right)= \left(QF\right)_{ZZ}, \qquad
\rho \left( G_{tt}-\frac{\mu_{tt}}{\mu}G \right)=  \left(QG\right)_{ZZ}, 
\en 
where $Q = Q(F^2+G^2, t)$. 
We assume that the Baker-Ericksen inequalities hold, so that $Q>0$ at all times \cite{Beatty}.
 
It is convenient to recast (\ref{equa}) as a single differential equation by introducing the complex quantity $\Lambda = F + \ii G$, to obtain
\be \label{complex1}
\rho \left(\Lambda_{tt}-\frac{\mu_{tt}}{\mu} \Lambda \right)= \left[Q \Lambda\right]_{ZZ},
\en
where now $Q = Q(\Lambda^2,t)$.

We seek solutions of (\ref{complex1}) in the form
\be \label{complex3}
\Lambda=\Omega(t) \exp\left[\ii \left(k Z+\phi(t) \right) \right],
\en
where the wave-number $k$ is a real constant, and $\Omega$ (the amplitude) and $\phi$ (the phase) are unknown real functions of time.
Notice that this type of solutions gives
\be \label{equi}
f(Z,t)=-\frac{\Omega(t)}{k} \sin\left(kZ+\phi(t)\right), 
\qquad g(Z,t)=\frac{\Omega(t)}{k} \cos\left(kZ+\phi(t)\right),
\en
for the shear motions, making the connection with the Carroll waves \eqref{carroll}: 
here we have the same spatial variations, and potentially richer temporal variations.

Now the full set of partial equations \eqref{complex1} reduces to the following system of ordinary differential equations,
\be \label{complex6}
\Omega''= \left[\phi'^2+\frac{\mu''}{\mu}-k^2\frac{Q}{\rho} \right] \Omega, 
\qquad \phi''\Omega+2 \Omega'\phi'=0,
\en
in the unknowns $\Omega(t)$ and $\phi(t)$ (now $Q = Q(\Omega^2,t)$). 
Carroll \cite{Ca1} derived this system in the unstretched case, when $\mu=\lambda \equiv 1$.

Direct integration of the second equation in (\ref{complex6}) gives  $\phi'=k_1/\Omega^2$,  where $k_1$ is an arbitrary constant, which we take to be positive without loss of generality. 
The amplitude $\Omega(t)$ is now the solution of the following nonlinear and {non-autonomous} ordinary differential equation,
\be \label{complex7}
\Omega''=  \left[\frac{\mu''}{\mu}-k^2\frac{Q}{\rho} \right] \Omega  + \frac{k_1^2}{\Omega^3}.
\en
The case $k_1=0$ (giving  $\phi(t)\equiv$ constant) has been discussed in detail by Pucci and Saccomandi \cite{Pucci}, who establish a clear relationship with Melde's problem of parametric resonance.

%%%%%%%%%%%%%%%%%%%%

\section{The Milne-Pinney Equation}

%%%%%%%%%%%%%%%%%%%%

We first specialize the equations to the Mooney-Rivlin solid, with strain-energy density
\be \label{MR}
W(I_1, I_2)= \textstyle{\frac{1}{2}} \mu_0 \left(\textstyle{\frac{1}{2}} + \beta\right)(I_1-3) + \textstyle{\frac{1}{2}}\mu_{0} \left(\textstyle{\frac{1}{2}} - \beta \right) (I_2-3),
\en
where $\mu_0>0$ is the infinitesimal shear modulus and $\beta$ is a constant such that $-1/2 \leq \beta \leq 1/2 $ (when $\beta=1/2$ we recover the special case of the neo-Hookean material). 
In that case, the generalized shear modulus is
\be
Q(t)=\mu_0 \left[\textstyle{\frac{1}{2}}+\beta + \left(\textstyle{\frac{1}{2}} - \beta\right)\lambda^{-1}(t)  \right],
\en
which is independent of $\Omega$,
and the governing equation   \eqref{complex7} becomes
\be \label{MR2}
\Omega'' - \left\{\frac{\mu''}{\mu}-\frac{\mu_0 k^2}{\rho} \left[\textstyle{\frac{1}{2}} + \beta+ \left(\textstyle{\frac{1}{2}} - \beta\right)\lambda^{-1} \right]\right\}\Omega -\dfrac{k_1^2}{\Omega^3}=0.
\en
This is indeed the Milne-Pinney or Ermakov equation. 
A detailed commentary about this equation is given in a recent paper by Leach and Andriopolous \cite{Leach}. 

Note that this is not the first time that this differential equation arises in the framework of nonlinear elastodynamics. 
In fact, it has arisen previously in the study of large amplitude oscillations of thin-walled tubes of Mooney-Rivlin materials. 
This problem was first treated by Knowles \cite{Knowles} for a general hyperelastic incompressible material, and the connection between this problem and the Milne-Pinney equation was made by Shahinpoor and  Nowinski \cite{Sha},  see also the book by Rogers and Ames \cite{RA}. 
Notice however that there is a noteworthy difference between the results in \cite{Sha} and the results here: the oscillations of a Mooney-Rivlin tube are governed by the Milne-Pinney equation in the approximation of a thin-walled tube, whereas here \textit{no} approximation at all was necessary to derive the equation. 
This fact was alluded to by Carroll in \cite{Ca1} who noticed the reduction \eqref{equi}, but did not use it to deduce some exact solution.    

To fix ideas, take  the following initial conditions 
\be \label{icOmega}
\Omega(0)=\Omega_0 >0, \qquad \Omega'(0)=0.
\en
Then the general solution of (\ref{MR2}) can be obtained by using two linearly independent solutions $u$ and $v$, say, of the linear equation
\be \label{aux}
\Omega'' - \left\{\frac{\mu''}{\mu} - \frac{\mu_0 k^2}{\rho} \left[\textstyle{\frac{1}{2}} + \beta + \left(\textstyle{\frac{1}{2}} - \beta\right)\lambda^{-1} \right] \right\}\Omega = 0,
\en
such that 
\be \label{ic}
u(0) = \Omega_0, \quad u'(0)=0, 
\qquad \text{and} 
\qquad
v(0) = 0, \quad v'(0) = 1/\Omega_0.
\en 
As shown by Pinney \cite{Pinney}, the general solution of (\ref{MR2}) is then
\be \label{pinney}
\Omega =\sqrt{u^2+ k^2_1 v^2}.
\en

An example can be worked out when $\lambda(t)$ is a \emph{constant, static equi-biaxial pre-stretch}, $\lambda =\lambda_S$, say. 
Then (\ref{MR2}) reduces to
\be \label{MR3}
\Omega'' + \omega^2_0 \Omega+\frac{k_1}{\Omega^3}=0,
\quad \text{where} \quad
\omega^2_0= \frac{\mu_0 k^2}{\rho} \left[\textstyle{\frac{1}{2}} + \beta + \left(\textstyle{\frac{1}{2}} - \beta\right)\lambda^{-1}_S \right]>0.
\en
In this case, $u=\cos(\omega_0 t)$, $v = \left[\sin (\omega_0 t)\right]/(\omega_0 \Omega_0)$, and the Pinney solution \eqref{pinney} is
\be \label{pinnesol}
\Omega(t)=\sqrt{\Omega_0^2 \cos^2(\omega_0 t)+\frac{k_1^2}{\omega^2_0 \Omega_0^2}\sin^2(\omega_0 t)}.
\en 
From the equation  $\phi'=k_1/\Omega^2$ we deduce the phase to be 
\be
\phi(t)= \tan^{-1}\left[ \dfrac{k_1}{\omega_0 \Omega_0^2} \tan(\omega_0 t)\right]+
\phi_0,
\en
where $\phi_0$ is the initial phase.

More complex solutions may be obtained by applying the approach of Shahinpoor and  Nowinski \cite{Sha}  to equation \eqref{MR2}.
For simplicity of exposition, we consider the neo-Hookean case, $\beta=1/2$. 
Then, the Milne-Pinney equation \eqref{MR2} reduces to
\be \label{MRNH}
\Omega'' - \left[\frac{\mu''}{\mu}-\frac{\mu_0 k^2}{\rho}\right]\Omega -\dfrac{k_1^2}{\Omega^3}=0.
\en
Now consider the case where the neo-Hookean solid is \emph{stretched linearly with time}, $\lambda(t)= 1 + a t$, say, where $a>0$ is a constant. 
Then $\mu''/\mu= 3 a^2/[4(1 + a t)^2]$. 
The following changes of function  and of  variable,
\be \label{change}
\omega = \dfrac{\Omega}{\Omega_0}  \qquad
\text{and} \qquad \zeta  = \alpha(1 + a t), 
\quad \text{where} \quad
\alpha \equiv \sqrt{\frac{\mu_0}{\rho}} \ \dfrac{k}{a},
\en
give a non-dimensional version of \eqref{MRNH},
\be \label{omega}
\omega_{\zeta \zeta} + \left(1 - \dfrac{3}{4 \zeta^2}\right)\omega - \dfrac{c}{\omega^3} = 0, \qquad \text{where} \qquad c \equiv \dfrac{\rho k_1^2}{\mu_0 k^2 \Omega_0^4},
\en
with initial conditions changed from \eqref{icOmega} to
\be \label{ic3}
\omega(\alpha) = 1, \qquad \omega'(\alpha) = 0.
\en
The linear Pinney equation corresponding to  \eqref{omega} is a Bessel equation, and we express the set of fundamental solutions in terms of Bessel functions of the first and second kind as
\be
u,v =  (\pi/4)\sqrt{\zeta/\alpha} \left[A_j J_1\left(\zeta\right) + B_j Y_1\left(\zeta\right) \right],
\qquad
j = 1,2,
\en
where
\begin{align}
& A_1 = -  Y_1(\alpha) + 2 \alpha Y_0(\alpha), 
&& A_2 = - 2 \alpha Y_1(\alpha), \notag \\
& B_1 =   J_1(\alpha) - 2 \alpha J_0(\alpha), 
&& B_2 = 2 \alpha J_1(\alpha).
\end{align}
These constants have been computed in order to accommodate the initial conditions for $u$ and $v$, which  are now:
\be
u(\alpha) = 1, \quad u'(\alpha)=0, 
\qquad \text{and} 
\qquad
v(\alpha) = 0, \quad v'(\alpha) = 1.
\en 

The resulting non-dimensional solution is $\omega = \left[ u^2 + c v^2\right]^{1/2}$. Its behavior depends on two constants: $c>0$ defined in \eqref{omega}, which is arbitrary because $k_1$ is arbitrary; and $\alpha > 0$ defined in \eqref{change}, which is equal to the ratio of the speed of infinitesimal shear waves to the speed of stretching. 
This connection leads us to define a `supersonic' range, $0<\alpha<1$ and a `subsonic' range, $\alpha>1$.

\begin{figure}
\begin{center}
\includegraphics*[width=7.9cm]{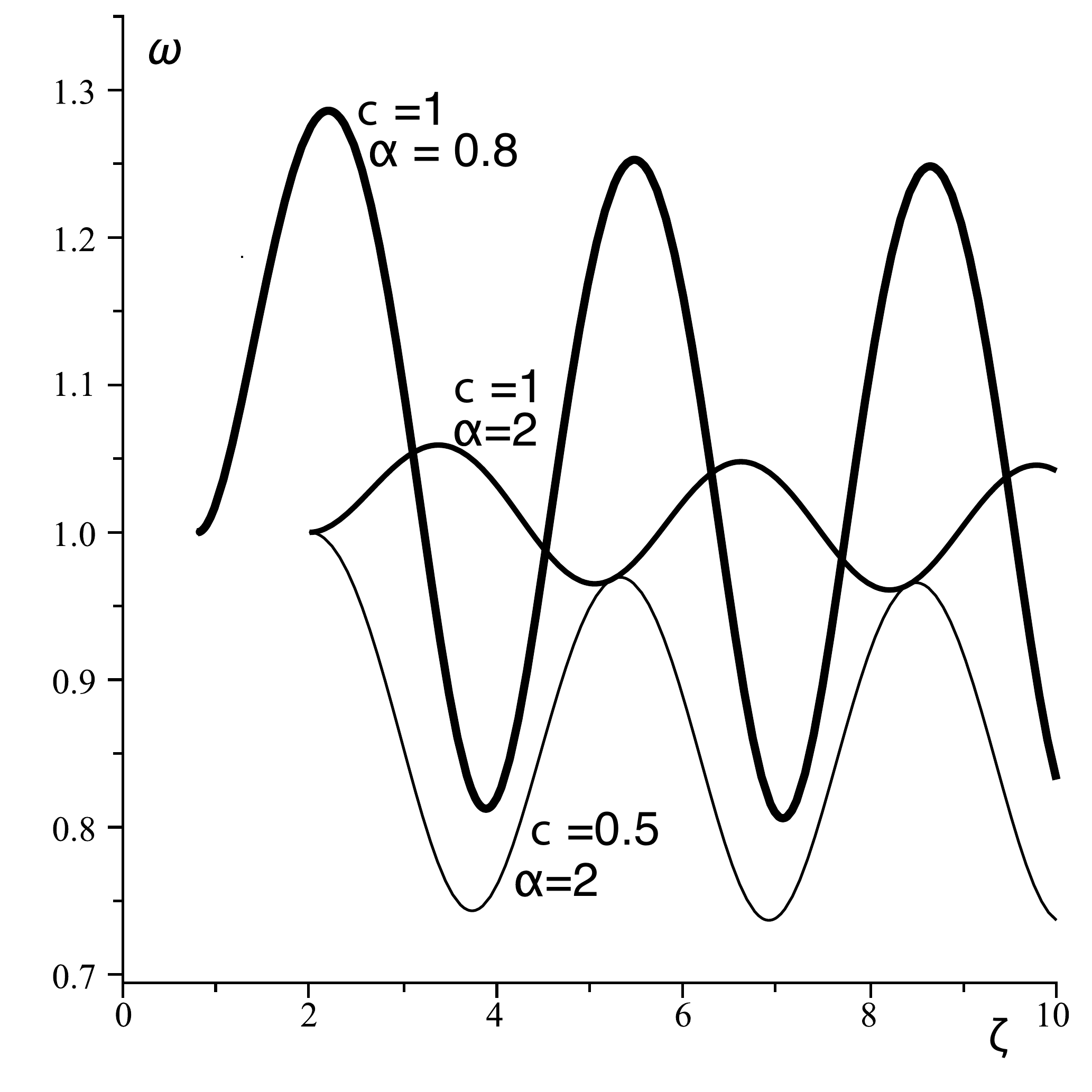}
\end{center}
\caption{Variations with time of the amplitude of a generalized Carroll wave in a neo-Hookean solid under a linearly growing stretch (non-dimensional units).
}
\label{fig_neo}
\end{figure}

At the origin $\zeta = \alpha$, we have \eqref{ic3}  and also $\omega_{\zeta \zeta} (\alpha) = c - 1 + 3/(4 \alpha^2)$.
Hence, when $\alpha$ is in the supersonic range $0<\alpha < \sqrt{3/4} \simeq 0.866$, the amplitude $\omega$ always grows at first because $\omega_\zeta (\alpha) =0$, $\omega_{\zeta \zeta}(\alpha) >0$. 
Otherwise, when $\alpha$ is in the supersonic range $\sqrt{3/4} < \alpha < 1$ or in the subsonic range $\alpha > 1$, the amplitude can either grow or decay, depending on the value of $c$ with respect to $ 1 - 3/(4 \alpha^2)$.
In Figure \ref{fig_neo} we show the early variations of $\omega$ when $c=1.0$, $\alpha= 0.8$ (thick plot), $c=1.0$, $\alpha = 2.0$ (medium thickness curve), and $c=0.5$, $\alpha = 2.0$ (thin curve).
In all cases, the amplitude oscillates and remains bounded over time.

%%%%%%%%%%%%%%%%%%%%%

\section{The Full Non-Linear Equation}

%%%%%%%%%%%%%%%%%%%%%

We established that the Milne-Pinney equation emerges from the specialization of the full non-linear equation \eqref{complex7} to the Mooney-Rivlin strain energy density \eqref{MR}. 
In fact, the exact solutions obtained subsequently by the Pinney method are valid for the whole class of \emph{general third-order  non-linear incompressible solids}. 
These are described by the following expansion of the strain energy $W$, 
\be \label{incomp3}
W = \mu_0 \ \text{tr}\left(\vec{E}^2\right)  + \frac{A}{3} \text{tr}\left(\vec{E}^3\right),
\en
where $\vec{E} = (\vec{F}^T \vec{F} - I)/2$ is the Green strain tensor and $A$ is a third-order elasticity constant in the notation of Hamilton et al. \cite{Hal}. 
Indeed, it is known since Rivlin and Saunders \cite{RiSa51} that when the Mooney-Rivlin strain energy \eqref{MR} is expressed in terms of $\text{tr}\left(\vec{E}^2\right)$ and $\text{tr}\left(\vec{E}^3\right)$, and neglecting higher powers, it embraces \eqref{incomp3}, once the following identification is made: $A = 2 \mu_0(2\beta - 3)$.
It follows that a solution valid for Mooney-Rivlin solids is also valid for all third-order incompressible solids (but the reverse is not true in general).

Further, we may use the exact Pinney solutions for Mooney-Rivlin solids as the starting point of a perturbation scheme in the \emph{fourth-order elasticity theory} of incompressible solids. 
As shown by Ogden \cite{Ogde74}, only three elastic constants are necessary to describe these.
In the notation of Hamilton et al. \cite{Hal}, the expansion of the strain energy density reads
\be \label{incomp4}
W = \mu_0 \ \text{tr}\left(\vec{E}^2\right)  + \frac{A}{3} \ \text{tr}\left(\vec{E}^3\right)   + D \ \left(\text{tr} (\vec{E}^2)\right)^2,
\en
where $A$ and $D$ are third-, and fourth-order elasticity constants, respectively.  
From Destrade et al. \cite{DeGM} we deduce that the following strain energy  
\be \label{MRD}
W= \mu_0 \left[ \left(\textstyle{\frac{1}{2}} + \beta\right)(I_1-3) + \left(\textstyle{\frac{1}{2}} - \beta \right) (I_2-3) + \frac{\gamma}{4} (I_1-3)^2 \right],
\en
covers \eqref{incomp4}, once terms of fifth-order in $\vec{E}$ and higher are neglected, and the following identifications apply,
\be
\beta = \dfrac{3}{2} + \dfrac{A}{4 \mu_0}, \qquad
\gamma = 1 + \dfrac{A/2+D}{\mu_0}.
\en

Using \eqref{MRD}, the corresponding generalized shear modulus $Q$ is easily computed as
\be
Q = \mu_0 \left[ \textstyle{\frac{1}{2}} + \beta + \left(\textstyle{\frac{1}{2}} - \beta \right) \lambda^{-1} + \gamma (2\mu^2 + \lambda^2 + \Omega^2-3) \right].
\en

As an example, consider a small-amplitude vibration of period $\omega$, superimposed upon a large static stretch $\lambda_S$, so that $\lambda(t)=\lambda_S [1 + \epsilon \cos(\omega t)]$, where $|\epsilon| << 1$. 
Then we expand  \eqref{complex7} up to term of order $\epsilon^1$.
We find
\begin{multline} \label{nl1}
\Omega'' + \omega_0^2 \Omega - \frac{k_1^2}{\Omega^3} = 
\frac{\mu_0 k^2}{\rho} \gamma (2 \lambda_S^{-1} + \lambda_S^2-3 - \Omega^2) \Omega 
\\ + \frac{\mu_0 k^2}{\rho} \epsilon \cos(\omega t) 
 \left[ \left(\textstyle{\frac{1}{2}} - \beta \right) \lambda^{-1}_S + 2 \gamma (\lambda_S^{-1} - \lambda_S^2)\right] \Omega,
 \end{multline}
 where $\omega_0$ is defined in \eqref{MR3}.

It is possible to consider several perturbation schemes that exploit the solution \eqref{pinnesol} as the zeroth-order approximation of equation \eqref{nl1}. 
The simplest of such perturbation scheme is obtained considering that the last term in \eqref{MRD} is a perturbation of the Mooney-Rivlin strain energy, i.e. that $\gamma = \hat \gamma \epsilon$, where $\hat \gamma = \mathcal{O}(1)$. 
With that assumption, we find that with
\be
\Omega'' + \omega_0^2 \Omega - \frac{k_1^2}{\Omega^3} = 
 \frac{\mu_0 k^2}{\rho} \epsilon  
 \left[ \left(\textstyle{\frac{1}{2}} - \beta \right) \lambda^{-1}_S \cos(\omega t)
  -  \hat \gamma (2 \lambda_S^{-1} + \lambda_S^2-3 - \Omega^2)\right] \Omega.
 \en
We may then perform a standard multiple scale expansion \cite{Stoker} starting from the exact solution \eqref{pinnesol} of the Milne-Pinney equation at $\epsilon=0$.  
In this way it is possible to investigate parametric resonance phenomena such as those revealed by the Melde string  experiment \cite{Melde}.
This has been done in the linear case by Lord Rayleigh \cite{Sound}, and more recently in a complete nonlinear context by Pucci and Saccomandi \cite{Pucci} (the main difference is that here we have traveling waves instead of standing waves.)
 
%%%%%%%%%%%%%%%%%%

\section{Concluding Remarks}

%%%%%%%%%%%%%%%%%%

The discovery of Carroll waves has had a long-lasting influence in the field of  wave propagation in nonlinear elasticity.
These solutions are elegant examples of exact reductions, and are also explicit examples of exact, closed-form solutions of an highly non-linear theory of solid mechanics. 
Moreover, the peculiarities of these solutions proved fundamental to a better understanding of the physical and mathematical properties
of the classical theory of elastodynamics. 
Fritz John \cite{John66} was the first to recognize this fact.
 
To give an idea of the recent research originating from the papers by Carroll \cite{Ca0, Ca1, Ca2, Ca3}, we mention: the note by Rubin and Rosenau \cite{Rubin} in the framework of the theory of a nonlinear string; the papers by Destrade and Saccomandi \cite{77, 97, 76} on rotating media, viscoelastic materials of differential type, and dispersive non-linear elastic materials, respectively; the paper by Rajagopal \cite{R1}; and the paper by Pucci and Saccomandi \cite{Pucci} on Melde's phenomena. 

The aim of this note was to point out a link between the Carroll waves and the Milne-Pinney equation. 
%It allowed us to find new exact solutions for Mooney-Rivlin and neo-Hookean materials, and to show to develop perturbation schemes for the investigation of the dynamics of a large class of non-linear elastic incompressible materials. 
It is clear that the solutions presented are somewhat puzzling from a mathematical point of view. 
Indeed, we know that global existence of body waves is guaranteed in nonlinear elasticity if the null condition is satisfied \cite{DOM}, but this result holds only for small initial data. 
The Carroll wave \eqref{carroll}, and its various generalizations, is an explicit example of global existence of solutions for any value of the amplitude of initial data. 
It is  still necessary the investigate the mathematical structure lying hidden behind such simple solutions.

%%%%%%%%%%%%%%%%%%%%%%%%%%%%%

\section*{Acknowledgments}

%%%%%%%%%%%%%%%%%%%%%%%%%%%%%

The work of MD is supported by a Marie Curie Fellowship for Career Development awarded by the Seventh Framework Programme of European Commission. The work of GS is partially supported by Italian GNFM. 

%%%%%%%%%%%%%%%%%%%%%%%%%%%%%

\end{document}